\documentstyle[aps,prb,twocolumn,epsf,floats]{revtex}
\begin{document}
\draft

\twocolumn[\hsize\textwidth\columnwidth\hsize\csname
@twocolumnfalse\endcsname

\title{Difference of optical conductivity between one- and two-dimensional
       doped nickelates}
\author{K. Tsutsui, W. Koshibae, and S. Maekawa}
\address{Institute for Materials Research, Tohoku University,
 Sendai 980-8577, Japan}
\date{October 20, 1998}
\maketitle
\begin{abstract}
We study the optical conductivity in doped nickelates, and
find the dramatic difference of the spectrum in the gap ($\omega$$\alt$4 eV)
between one- (1D) and two-dimensional (2D) nickelates.
The difference is shown to be caused by the dependence of hopping integral on
 dimensionality.
The theoretical results explain consistently the experimental data in 1D and
 2D nickelates, Y$_{2-x}$Ca$_x$BaNiO$_5$ and La$_{2-x}$Sr$_x$NiO$_4$,
respectively.
The relation between the spectrum in the X-ray aborption experiments and the
optical conductivity in La$_{2-x}$Sr$_x$NiO$_4$ is discussed.
\end{abstract}


]
\narrowtext

Nickelates have received special attention since the discovery of high $T_c$
cuprates.
An antiferromagnetic insulator La$_2$NiO$_4$ has the same layered structure as
that of one of the parent compounds of high $T_c$ superconductors,
La$_2$CuO$_4$.
In contrast with the cuprate with spin-1/2, the nickelate is a spin-1 system.
Therefore, the hole-doped nickelate, La$_{2-x}$Sr$_x$NiO$_4$, provides an
opportunity to examine a role of the spin background in the carrier dynamics
in two-dimension.
Another doped nickelate, Y$_{2-x}$Ca$_x$BaNiO$_5$, has also been studied
extensively.\cite{ditusa94,ito98}
In the undoped one, Y$_2$BaNiO$_5$, apex-linked NiO$_6$ octahedra are arranged
in one-dimension and exhibit a spin-1 chain.

The spectrum of the optical conductivity in the low energy region
 ($\omega$$\alt$4 eV) reflects the key parameters for the electronic structure 
such as Coulomb interaction, Hund's-rule coupling, superexchange interaction, 
hopping integral of carriers, electron-phonon interaction and so on.  
The experiments in one-dimensional (1D) nickelate, Y$_{2-x}$Ca$_x$BaNiO$_5$,
and two-dimensional (2D) one, La$_{2-x}$Sr$_x$NiO$_4$, have been performed by 
several groups.\cite{ito98,bi90,bi93l,bi93b,ido91,katsufuji96}
Both the undoped nickelates are insulators with the charge excitation gap of
about 4 eV.  
Upon doping of holes, the spectrum appears in the gap.  
There are dramatic differences in the spectrum between 1D and 2D nickelates.
In 2D, the spectrum spreads in the gap ($\omega$$\alt$4 eV)
and is broader than that in 1D.
At around 1 eV, the spectrum shows two-peak structure in 2D,
whereas it has a single peak in 1D.

In this paper, we propose that such a remarkable difference between 1D and 2D
nickelates is caused by the hopping integral of $3d$ electrons.
The hopping integral exists between all of the neighboring $e_g$ orbitals in
 2D.
In 1D, on the other hand, the integral is restricted within the
 $3d(3z^2$$-$$r^2)$ orbitals, where the chain direction is taken to be the
 $z$-axis.
It is shown that the dependence of the hopping integral on dimensionality
 affects the dynamics of carriers and the spectrum in optical conductivity.
We examine the optical conductivity in 1D and 2D nickelates by using
the numerically exact diagonalization method.  
We find the dramatic difference of the spectrum between 1D and 2D:
The spectrum in 2D spreads in the gap in contrast with that in 1D.
In 2D, the two-peak structure is obtained in the low energy region.  
One of the peaks is due to a low-spin state induced by a doped carrier and 
is governed by the Hund's-rule coupling.  
The other originates in the antiferromagnetic spin background disturbed by the
carriers.
In 1D, the peak for the low spin state is strongly suppressed.  
The difference of the spectrum is well described by the dependence of hopping
integral on dimensionality.  
We also examine the relation between the optical conductivity and
the single-particle excitation spectrum.
The experimental data of the optical conductivity and X-ray absorption
spectroscopy are discussed in the light of theoretical results.  

A Ni$^{2+}$ ion in a NiO$_6$ octahedron has two electrons in the $e_g$
orbitals, $3d(x^2$$-$$y^2)$ and $3d(3z^2$$-$$r^2)$.
Our starting model consists of the following terms;
hopping integrals between the $e_g$ orbitals on nearest-neighbor sites, 
on-site Coulomb interactions and Hund's-rule coupling between electrons
in the $e_g$ orbitals.  The Hamiltonian is written as, 
\begin{equation}
\label{ham1}
   H = H_t + H_{\rm I}.
\end{equation}
The kinetic energy of $3d$ electrons $H_t$ is written as,
\begin{equation}
H_t = \sum_{\langle l,m\rangle,\sigma,\mu,\nu}t_{lm}^{\mu \nu}
      (c_{l,\mu,\sigma}^{\dag}c_{m,\nu,\sigma} + {\rm h.c.}), 
\label{hamt}
\end{equation}
where $c_{l,\mu,\sigma}^{\dag}$ is the creation operator of electron with
spin $\sigma$ in $\mu$ orbital at site $l$, and $t_{lm}^{\mu \nu}$ is the
hopping integral between orbital $\mu$ at site $l$ and $\nu$ at $m$.
The integral $t_{lm}^{\mu \nu}$ depends crucially on the dimensionality of
the system and is expressed as, 
\begin{eqnarray}
\left(\begin{array}{cc}
 -t/3        &  \pm t/\sqrt3\\
\pm t/\sqrt3 &     -t 
\end{array}\right),
\label{hamt2d}
\end{eqnarray}
in the $xy$ plane in 2D and,
\begin{eqnarray}
\label{hamt1d}
\left(\begin{array}{cc}
-4t/3     &  0    \\
    0     &  0
\end{array}\right),
\end{eqnarray}
in 1D, where the chain direction is taken to be $z$-axis.  
Here, $t$ is the integral between neighboring $3d(x^2$$-$$y^2)$ orbitals and
the $+(-)$ sign in the off-diagonal terms in Eq.~(\ref{hamt2d})
corresponds to the integral in the $x(y)$ direction in 2D.
In 2D, electrons in both $e_g$ orbitals contribute to the transport.  
In 1D, on the other hand, an electron in the $3d(x^2$$-$$y^2)$ orbital is
localized in each Ni ion since only the integral between neighboring
$3d(3z^2$$-$$r^2)$ orbitals is finite.  
As will be shown below, the dependence of the hopping integral on
dimensionality causes a crucial role in the electronic structure.

The electron-electron interaction $H_{\rm I}$ is given by,
\begin{eqnarray}
\label{hami}
H_{\rm I}&=&U\sum_{l,\mu} n_{l,\mu,\uparrow}n_{l,\mu,\downarrow}
            + U'\sum_l n_{l,u}n_{l,v} \nonumber \\
         &-&2K\sum_l 
       (\vec{S}_{l,u} \cdot \vec{S}_{l,v}+{1\over4}n_{l,u}n_{l,v}) \nonumber\\
         &+&K\sum_l
       (c_{l,u,\uparrow}^\dagger c_{l,u,\downarrow}^\dagger 
        c_{l,v,\downarrow}c_{l,v,\uparrow}+{\rm h.c.}),
\end{eqnarray}
where $u$ and $v$ denote the $3d(3z^2$$-$$r^2)$ and $3d(x^2$$-$$y^2)$ orbitals,
respectively, $n_{l,\mu,\sigma}$$=$$c_{l,\mu,\sigma}^\dagger c_{l,\mu,\sigma}$,
$n_{l,\mu}$$=$$n_{l,\mu,\uparrow}$$+$$n_{l,\mu,\downarrow}$, $U$ and $U'$ are
the intra- and inter-orbital Coulomb interactions, and $K$ is the Hund's-rule
coupling with $U$$=$$U'$$+$$2K$.

The O $2p$ orbitals are not taken into account explicitly, since we are
 interested in the dependence of optical conductivity on dimensionality.
When the effect of the O $2p$ orbitals is estimated in the NiO$_6$ cluster
model, we find that the atomic parameters are reduced by $\sim$30\%.  From
the observed atomic parameters in La$_2$NiO$_4$,
\cite{kuiper91,eisaki92,pellegrin96,kuiper98}
the values of $U$ and $K$ are taken to be $\sim$5 and $\sim$0.7 eV,
respectively, and the value of $t$ is estimated to be
 $t$$\sim$$T_{pd}^2/\Delta$$\sim$10$^{-1}$ eV.
Since the observed crystal field splitting of $e_g$ orbitals in La$_2$NiO$_4$
is much smaller than the Coulomb interaction,\cite{kuiper98} we neglect it
in this study.  
The importance of phonons in the low energy region ($\omega$$\alt$0.5 eV)
has been discussed.\cite{bi90,bi93l,bi93b,ido91,katsufuji96}
Since we are interested in the higher energy region, 
the electron-phonon interaction is neglected in the present study.

We calculate optical conductivity and single-particle excitation spectra 
in the numerically exact diagonalization method on 8 site clusters in 1D and 
$\sqrt{8}$$\times$$\sqrt{8}$ site clusters in 2D with periodic boundary
condition.
The values of $U/t$ and $K/t$ used in the calculation are 14 and 2,
respectively.
The optical conductivity is written as,
\begin{equation}
\label{opt}
\sigma(\omega)= D\delta(\omega) + \frac{\pi}{N_S} \sum_{n\neq 0}
                \frac{|\langle n|j_a|0 \rangle|^2}{E^N_n-E^N_0}
                \delta(\omega-E^N_n+E^N_0),
\end{equation}
where $|0\rangle$ and $|n\rangle$ are the ground state and $n$-th
excited state with energies $E^N_0$ and $E^N_n$, respectively, $D$ denotes
the Drude constant,
and $N_S$ and $N$ are the system size and number of electrons, respectively.
The Drude weight is estimated from the sum rule.\cite{maldague77,stephan90}
The $a$-axis component of the current operator $j_a$ is given by,
\begin{equation}
\label{current}
j_a = -i\sum_{l,\mu,\nu,\sigma}t_{l+a,l}^{\mu \nu}
      (c_{l+a,\mu,\sigma}^{\dag}c_{l,\nu,\sigma} - {\rm h.c.}),
\end{equation}
where $a$$=$$z$ and $x$ in 1D and 2D, respectively.
In the same way, the single particle excitation spectrum is expressed as,
\begin{eqnarray}
\label{akw}
A_\mu(k,w) &=& \sum_n |\langle n|c_{k,\mu,\sigma}|0\rangle|^2
               \delta(-\omega-E_n^{N-1}+E_0^N) \nonumber \\
           &+& \sum_n |\langle n|c_{k,\mu,\sigma}^\dagger|0\rangle|^2
               \delta(\omega-E_n^{N+1}+E_0^N),
\end{eqnarray}
where $c_{k,\mu,\sigma}$ is the Fourier transform of $c_{l,\mu,\sigma}$.

\begin{figure}
\epsfxsize=8cm
\centerline{\epsffile{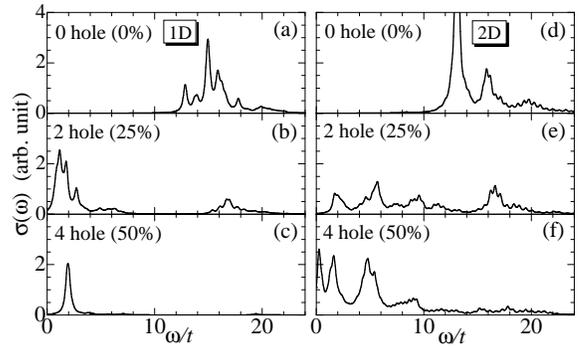}}
\caption{
Optical conductivity $\sigma(\omega)$ in 1D and 2D.
(a), (d): non-hole doping, (b), (e): two hole (25\% hole) doping, and
(c), (f): four hole (50\% hole) doping.
The parameters $U/t$$=$14, $K/t$$=$2 are used.  
The $\delta$-functions are convoluted with a Lorentzian broadening of 0.2$t$.
The Drude constants ($D$) are not displayed.
The values of $D$ are; (a):0.00, (b):2.96, (c):6.68, (d):$-$1.27, (e):3.10,
and (f):2.50 in the same unit.
}
\label{fig1}
\end{figure}

In Fig.~\ref{fig1}, the numerical results of $\sigma(\omega)$ in 1D and 2D
with no-hole (insulating), two holes (25\% doping)
and four holes (50\% doping) are shown.
In the undoped states, a charge excitation gap exists at $\omega/t$$\simeq$12.
When the value of $t$ is taken to be $\sim$0.33 eV, the gap energy is
estimated to be $\sim$4 eV, which is consistent with the experimental
data.\cite{ito98,ido91}
Upon doping of holes, the spectrum appears in the gap.\cite{drude}
The shape of the spectrum is quite different between 1D and 2D systems.
In 2D, the spectrum has some characteristic peaks in the gap
($\omega/t$$\alt$12), whereas in 1D it is rather sharp in
the low energy region($\omega/t$$\simeq$2).

To clarify the nature of the spectrum in the gap, let us first consider
the charge excitations in the strong coupling limit($t$$\ll$$U,K$).
In the undoped states, each site has two electrons in the $e_g$ orbitals with
high spin ($S$$=$1) configuration due to the Hund's-rule coupling.
Therefore, the charge excitation gap has the energy $\sim$$U$.
In the hole-doped systems, there exist singly occupied sites with $S$$=$1/2.
The hopping of an electron from a doubly occupied ($S$$=$$1$) site to a
singly occupied ($S$$=$1/2) site costs the energy less than the gap energy.
After the hopping, the high spin ($S$$=$1) or low spin ($S$$=$0) state appears 
at the doubly occupied site.  
In the former case, the excitation is caused by the hopping of
electron which disturbs the spin background.
In the latter case, on the other hand, the excitation is characterized
by the energy scale with $K$.
These excitations appear in the gap of $\sigma(\omega)$.

\begin{figure}
\epsfxsize=8cm
\centerline{\epsffile{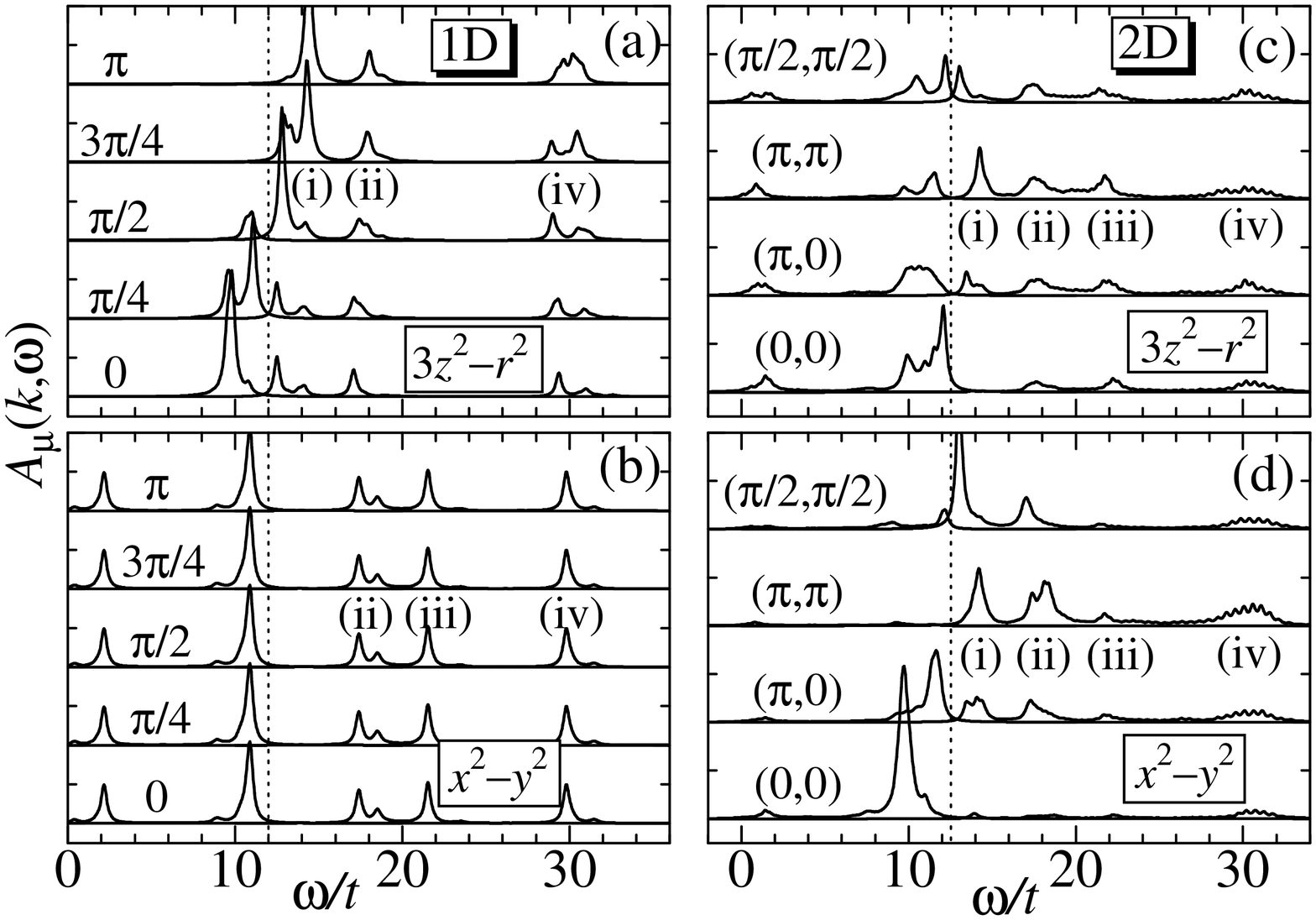}}
\caption{
Single particle excitation spectra $A_\mu(k,\omega)$ with four holes
(50\% hole doping case).
Left and right panels are for 1D and 2D, respectively.
Upper and lower panels are the spectra for electrons in
$3d(3z^2$$-$$r^2)$ and $3d(x^2$$-$$y^2)$ orbitals, respectively.
The dotted lines denote Fermi level.
The $\delta$-functions are convoluted with a Lorentzian broadening of 0.4$t$.
The parameter values are the same with those in Fig.~1.  
}
\label{fig2}
\end{figure}

Let us next calculate the single particle excitation spectrum
$A_\mu(k,\omega)$ and examine the relation between $A_\mu(k,\omega)$ and
$\sigma(\omega)$.
Figure~\ref{fig2} shows $A_\mu(k,\omega)$ for 50\% hole doped systems
in 1D and 2D.
The peaks (i)$\sim$(iii) are characterized by the $^3A_2$ (high spin),
$^1E$ (low spin), and $^1A_1$ (low spin) states at the site
where an electron is added, respectively, and the peak (iv) is given by
the triply occupied states at the site.
In Fig.~\ref{fig2}(a), the peak (iii) does not exist,
because the $^1A_1$ state, in which two electrons are in the same $e_g$
orbital, can not be created by adding an electron in $3d(3z^2$$-$$r^2)$
orbital.\cite{hole1d}
In Fig.~\ref{fig2}(b), the spectrum is independent of $k$,
since the hopping matrix elements for $3d(x^2$$-$$y^2)$ orbitals do not
exist in 1D and electrons in the orbitals are localized in Ni ions.
In Figs.~\ref{fig2}(c) and (d), the hopping matrix elements for both orbitals
cause the broadening of the spectra of (ii)$\sim$(iv).

\begin{figure}
\epsfxsize=8cm
\centerline{\epsffile{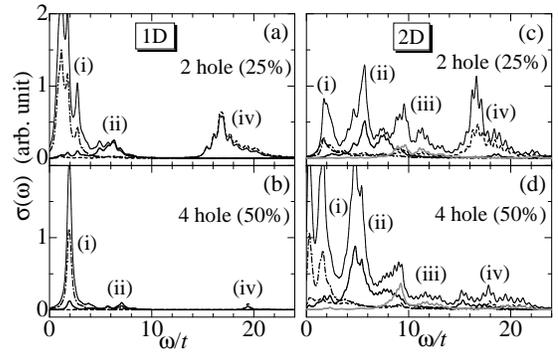}}
\caption{
Optical conductivity are analyzed by using the
operators given by Eq.~(\ref{cup}).
Left and right panels are for 1D and 2D, respectively.
(a), (c): two hole (25\% hole) doping, (b), (d): four hole (50\% hole) doping.
The thin lines are the same results shown in Fig.~\ref{fig1}.
The dot-dashed, thick, gray, and dashed lines denote spectra for
$\alpha$$=$$^3A_2$, $^1E$, $^1A_1$, and 
the triply occupied state, respectively.
The $\delta$-functions are convoluted with a Lorentzian broadening of 0.2$t$.
The parameter values are the same with those in Fig.~1.  
}
\label{fig3}
\end{figure}

In 1D, since only $3d(3z^2$$-$$r^2)$ orbital contributes to the current,
the excitations from the states below the Fermi level to those of (i) and
(ii) in Fig.~\ref{fig2}(a) appear in the gap of $\sigma(\omega)$.
On the other hand, in 2D, the excitations to the states of (i),
(ii), and (iii) in both Figs.~\ref{fig2}(c) and (d) appear in the gap,
so that the spectrum is broader in 2D than that in 1D.
To resolve these excitaions in the optical conductivity, we introduce the
operator $\tilde{j}_a$ given by,
\begin{equation}
\label{cup}
            \tilde{j}_a = -i\sum_{l,\mu,\nu,\sigma}t_{l+a,l}^{\mu \nu}
                         P_{l,\alpha}c_{l,\mu,\sigma}^\dagger
                         (c_{l-a,\nu,\sigma} - c_{l+a,\nu,\sigma}),
\end{equation}
where $P_{l,\alpha}$ is the projection operator which restricts the state
to $\alpha$ at site $l$. 
For example, the projection operator for $\alpha$$=$$^1A_1$ with two electrons
is written as,
\begin{eqnarray}
\label{prj}
P_{l\ ^1A_1}&=&\frac{1}{2}\{
      n_{l,u,\uparrow}           n_{l,u,\downarrow}
   (1-n_{l,v,\uparrow})       (1-n_{l,v,\downarrow})        \nonumber \\
&+&   n_{l,v,\uparrow}           n_{l,v,\downarrow}
   (1-n_{l,u,\uparrow})       (1-n_{l,u,\downarrow})        \nonumber \\
&+&   c_{l,u,\uparrow}^\dagger   c_{l,u,\downarrow}^\dagger
      c_{l,v,\downarrow}         c_{l,v,\uparrow}           
+   c_{l,v,\uparrow}^\dagger   c_{l,v,\downarrow}^\dagger
      c_{l,u,\downarrow}         c_{l,u,\uparrow}\}         .
\end{eqnarray}
We calculate the spectrum Eq.~(\ref{opt}) by replacing $j_a$ with
$\tilde{j}_a$.
The results for $\alpha$$=$$^3A_2$, $^1E$, $^1A_1$, and triply occupied state
are shown in Fig.~\ref{fig3}.
The dot-dashed, thick, gray, and dashed lines denote the spectra for
$\alpha$$=$$^3A_2$,
$^1E$, $^1A_1$, and the triply occupied state, respectively.
The thin lines show the same results as those in Fig.~\ref{fig1}.
As shown in Fig.~\ref{fig3}, the peaks (i), (ii), and (iii) are well described 
by the final states with $\alpha$$=$$^3A_2$, $^1E$, and $^1A_1$, respectively.

The spectra of (i) in both 1D and 2D are brought about by the
motion of carriers which disturbs the spin configuration.
They are similar to the mid-gap spectra in high $T_c$ cuprates.
Note, however, that the spectra exist even in 1D $S$$=$$1$ systems.
This is in contrast with that in 1D $S$$=$1/2 systems
where such spectra do not exist.\cite{stephan90}

\begin{figure}
\epsfxsize=6.5cm
\centerline{\epsffile{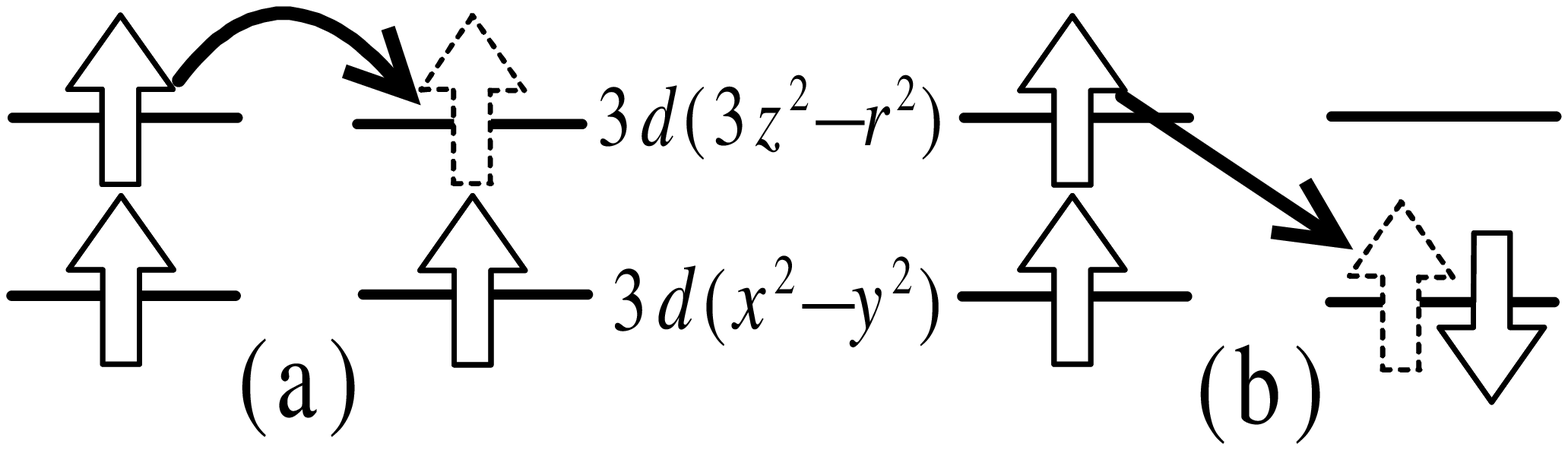}}
\caption{
The difference of the hopping matrix elements between 1D and 2D.
(a) In 1D, the hopping matrix elements are restricted within
$3d(3z^2$$-$$r^2)$ orbitals.
(b) In 2D, the off-diagonal hopping matrix elements induce the low spin state.
}
\label{fig4}
\end{figure}

The spectra of (ii) and (iii) are due to the excitations
in which the motion of carriers changes doubly occupied sites from high spin
to low spin states.
These spectra in 1D are smaller in magnitude than those in 2D.
In particular, the spectrum of (iii) does not exist in 1D,
because the hopping matrix element for $3d(x^2$$-$$y^2)$ orbitals does not
exist and the $^1A_1$ state is excluded in the optical process.
The small magnitude for the spectrum of (ii) in 1D is understood as follows;
since the hopping matrix element is restricted within the $3d(3z^2$$-$$r^2)$
orbitals in 1D, the ferromagnetic spin alignment is stabilized around a singly
occupied site due to the double exchange interaction (see Fig.~\ref{fig4}(a)).
In 2D, on the other hand, the off-diagonal hopping matrix elements
induce the antiferromagnetic superexchange interaction and reduce the
stability of ferromagnetic spin alignment (see Fig.~\ref{fig4}(b)).
Therefore, the probability for the excitation to the low spin state is
smaller in 1D than that in 2D, and the intensity of the peak (ii) is
reduced in 1D.

Experimentally, it has been shown that the spetrum of the optical conductivity
in La$_{2-x}$Sr$_x$NiO$_4$ spreads in the gap and has two peaks at 0.6 and
1.5 eV.\cite{ido91}  The two peaks grow with doping.
Figs.~\ref{fig3}(c) and (d) explain this fact; two peaks (i) and (ii) grow
with doping, and the energies are $\sim$2$t$ and $\sim$5$t$ which are
estimated to be $\sim$0.6 and $\sim$1.5 eV, respectively, where
$t$$\sim$0.3 eV is taken.\cite{polaron}
The appearance of the spectrum of (iii) is also consistent with the
observation that the spectrum spreads in the gap.  
In 1D, the peak (i) in Fig.~\ref{fig3}(a) is much larger than (ii),
and the peak (i) explains well the single peak structure
in Y$_{2-x}$Ca$_x$BaNiO$_5$.\cite{ito98}

Let us discuss the O $1s$ near-edge X-ray absorption spectroscopy(XAS).
Pellegrin {\it et al.}\cite{pellegrin96} have observed that there appear the
peaks at 528.7 and 530 eV in the polarized XAS upon doping of holes.
They identified the two peaks as the high spin state and the low spin one,
respectively.
In our theory, the states correspond to the states of (i) and (ii)
in $A_v(k,\omega)$.\cite{tetra}
It is shown that the excitations (i)$\sim$(iv) in $A_\mu(k,\omega)$ 
bring about the peaks (i)$\sim$(iv) in $\sigma(\omega)$, respectively.  
Thus, we find that the peaks at 0.6 eV and 1.5 eV in the optical
conductivity in La$_{2-x}$Sr$_x$NiO$_4$ correspond to the
excitations at 528.7 and 530 eV in XAS, respectively.

In summary, we have examined the optical conductivity in 1D and 2D
nickelates and found that the dramatic difference occurs between
them, since the hopping integral of $3d$ electrons between neighboring Ni sites
has the off-diagonal term in the $e_g$ orbitals in 2D nickelates,
whereas the integral is restricted within the $3d(3z^2$$-$$r^2)$ orbitals in
1D ones.
The theoretical results explain consistently the experimental data in
Y$_{2-x}$Ca$_x$BaNiO$_5$ and La$_{2-x}$Sr$_x$NiO$_4$.
We have also shown the relation between the spectra in XAS and optical
conductivity in La$_{2-x}$Sr$_x$NiO$_4$.

Authors thank S. Uchida, H. Eisaki, and T. Ito for giving us the experimental
data prior to publication, and S. Ishihara for valuable discussion. 
This work was supported by Priority-Areas Grants from the Ministry 
of Education, Science, Culture and Sport of Japan, CREST, and NEDO.
Computations were carried out in ISSP, Univ. of Tokyo, IMR, Tohoku Univ.,
and Tohoku Univ.


\begin{references}
\bibitem{drude}
In the doped system, the numerical calculation gives the finite Drude weight
in contrast with experiment.  The origin to suppress the weight in the real
materials is, at present, not known.  However, if random potentials cause
the insulating state, we can show that the qualitative change does not occur
in the gap.
\bibitem{hole1d}
In 1D, we find that holes are always introduced into $3d(3z^2$$-$$r^2)$
orbitals in the ground state in lightly doped case (up to 50\% hole doping).
Thus, the number of the electrons in each $3d(x^2$$-$$y^2)$ orbital is
exactly 1.
\bibitem{polaron}
The peak at $\sim$0.6 eV in La$_{2-x}$Sr$_x$NiO$_4$ may include the small
polaron effect.\cite{bi90,bi93l,bi93b}
\bibitem{tetra}
The distortion of NiO$_6$ octahedron induces the crystal field splitting 
($E_z$) between $e_g$ orbitals and the reduction of the symmetry
$^1E$$+$$^1A_1$ (in $O_h$ symmetry) $\rightarrow$ 2$^1A_1$$+$$^1B_1$
(in $D_{4h}$ symmetry).  Two peaks are observed in the XAS,\cite{pellegrin96}
which correspond to the $^1B_1$ and high spin states.  However, the energy
difference between them is not modified by $E_z$, since each of
$3d(3z^2$$-$$r^2)$ and $3d(x^2$$-$$y^2)$ orbitals is occupied by an electron
in the $^1B_1$ and high spin states.
\end{references}
\end{document}